\documentclass[aps,prl,superscriptaddress,twocolumn,showpacs]{revtex4}

\usepackage{amsmath,epsfig,color,amssymb}
\usepackage{graphicx}
\usepackage{graphicx}
\usepackage{dcolumn}
\usepackage{bm}
\usepackage{graphicx}
\usepackage{amsmath, amsfonts, amssymb,mathrsfs}
\usepackage{pstricks}
\usepackage{amsxtra}
\usepackage{amsthm}
\usepackage{natbib}
\usepackage{color}

\makeatletter
\@ifundefined{textcolor}{}
{%
 \definecolor{BLACK}{gray}{0}
 \definecolor{WHITE}{gray}{1}
 \definecolor{RED}{rgb}{1,0,0}
 \definecolor{GREEN}{rgb}{0,1,0}
 \definecolor{BLUE}{rgb}{0,0,1}
 \definecolor{CYAN}{cmyk}{1,0,0,0}
 \definecolor{MAGENTA}{cmyk}{0,1,0,0}
 \definecolor{YELLOW}{cmyk}{0,0,1,0}
}

\def\be{\begin{equation}}
\def\ee{\end{equation}}
\def\bea{\begin{eqnarray}}
\def\eea{\end{eqnarray}}
\newcommand{\ket}[1]{\mbox{$|#1\rangle$}}
\newcommand{\bra}[1]{\mbox{$\langle#1|$}}

\def\be{\begin{equation}}      
\def\ee{\end{equation}}
\def\beu{\begin{equation*}}   
\def\eeu{\end{equation*}}

\providecommand{\abs}[1]{\left\lvert#1\right\rvert}   
\providecommand{\ket}[1]{\left|#1\right\rangle}

\providecommand{\bra}[1]{\left\langle#1\right|}
\providecommand{\mean}[1]{\left\langle#1\right\rangle}

\definecolor{new}{rgb}{.08,.05,.8}

\newcommand{\delete}[1]{{}}

\graphicspath{}

\begin{document}
\title{Phonon-Assisted Gain in a Semiconductor Double Quantum Dot Maser}
\author{M.~J.~Gullans}
\affiliation{Joint Quantum Institute, National Institute of Standards and Technology, Gaithersburg, MD 20899, USA}
\affiliation{Joint Center for Quantum Information and Computer Science, University of Maryland, College Park, MD 20742, USA}
\author{Y.-Y. Liu}
\author{J. Stehlik}
\affiliation{Department of Physics, Princeton University, Princeton, NJ 08544, USA}
\author{J. R. Petta}
\affiliation{Department of Physics, Princeton University, Princeton, NJ 08544, USA}
\affiliation{Department of Physics, University of California, Santa Barbara, CA 93106, USA}
\author{J.~M.~Taylor}
\affiliation{Joint Quantum Institute, National Institute of Standards and Technology, Gaithersburg, MD 20899, USA}
\affiliation{Joint Center for Quantum Information and Computer Science, University of Maryland, College Park, MD 20742, USA}

\date{\today}
\begin{abstract}
We develop a microscopic model for the recently demonstrated double quantum dot  (DQD) maser. In characterizing the gain of this device we find that, in addition to the direct stimulated emission of photons, there is a large contribution from the simultaneous emission of a photon and a phonon, i.e., the phonon sideband. We show that this phonon-assisted gain typically dominates the overall gain which leads to masing. Recent experimental data are well fit with our model.
\end{abstract}
\pacs{73.21.La, 42.50.Pq, 78.67.Hc, 85.35.Gv}
\maketitle


The coherent generation of light in a laser provides fundamental insights into the interaction between light and matter \cite{LaserBook}.  Lasers operating in the few-emitter limit probe this interaction at the level where quantum effects are crucial for understanding the device operation \cite{Mu92,Rice94,Bjork94}.   Single emitter lasers were first demonstrated in atomic systems \cite{McKeever03,Walther06} and, subsequently, extended to solid-state systems, where one must contend with a strong coupling of the emitter to the surrounding environment \cite{Xie07,Nomura10,Nakamura07,Chen14}.  

Several groups have explored the possibility of achieving a maser with gate-defined semiconductor quantum dots as the gain medium \cite{Childress04,Jin11,Kulkarni14,Liu14a}.  Recently a successful demonstration of such a maser was achieved by coupling two InAs nanowire double quantum dots (DQDs) to a microwave cavity \cite{Liu14b}.  Due to the large Coulomb charging energy $E_c \sim 5$ meV, 
these systems provide tunable gain from GHz to THz frequencies using external gate voltages. 
Operating in the few-emitter limit, they may enable the creation of quantum states of light  \cite{MeystreBook} and
 entangled states of DQDs and light \cite{Kimble08,Petersson12,Bassett13,Deng14b}.
 The strong environmental coupling in these devices allows the study of competing emission mechanisms, e.g., phonon versus photon \cite{Bergenfeldt13}.  While the role of electron-phonon coupling has been considered in previous work on optical quantum dot lasers \cite{Nielsen04,Wijnen08,Majumdar11,Quilter15}, electrically driven quantum dots probe a much lower energy scale. 
 Finally, previous theoretical work predicts a small, narrow gain feature in the DQD emission spectrum
 \cite{Childress04,Jin11,Kulkarni14}. This is in contrast with the experimental results, where high gain is observed over a much larger energy range \cite{Liu14a,Liu14b}.  Resolving this discrepancy is crucial for future applications of the DQD-cavity system to both maser operation and quantum information tasks.

In this Letter, we develop a microscopic model for the recently demonstrated DQD maser \cite{Liu14b}.  In characterizing the gain of this device, we find, in addition to the direct stimulated emission of photons into the cavity, a large contribution from transitions that involve the simultaneous emission of a photon and a phonon, i.e., the phonon sideband. These effects have not been considered in  previous related work \cite{Childress04,Jin11,Kulkarni14}.  
Under typical experimental conditions, the phonon sideband dominates the gain and, therefore, sets the energy range over which masing occurs.
We find the experimental data from Ref.\ \cite{Liu14b} are well fit with  a theoretical model accounting for this phonon process.

A schematic of a DQD maser is shown in Fig.\ 1.   The gain medium consists of one or several DQDs coupled to the common mode of a microwave resonator [Fig.\ 1(a)].  With a bias applied across the DQDs, current flows via single electron tunneling and, in Ref.\ \cite{Liu14a,Liu14b}, gain was observed in the cavity transmission. However, 
this gain occurred over a much wider range of DQD transition frequencies than the cavity resonance and was much larger than predicted from a 
Jaynes-Cummings model.  We can understand the broadening of the gain at a qualitative level by noting that the electron-phonon interaction will dress the electronic states of the quantum dot with the phonons in the nanowire.  This leads to a phonon sideband whereby energy is conserved through the simultaneous emission of a phonon and cavity photon, as illustrated in Fig.\ 1(b-c).   To understand how this affects the maser note that effective maser operation requires a large photon emission rate, a large population inversion, and a rapid repumping rate.  
The peak emission rate for the direct process occurs when the DQD is on resonance with the cavity.  
Without precise tuning of the system, this does not always correspond to the optimal operating point for the maser (e.g., as is the case in Ref.\ \cite{Liu14a,Liu14b}).  Furthermore, in the presence of charge noise it is difficult to stabilize the DQD at the resonance condition.  In this far-off resonant regime, we show that the phonon sideband strongly dominates over the gain from direct photon emission.  
As a result, the DQD maser dynamics is typically dominated by this phonon-assisted process and not direct photon emission.


 \begin{figure}[t]
\begin{center}
\includegraphics[width=.49 \textwidth]{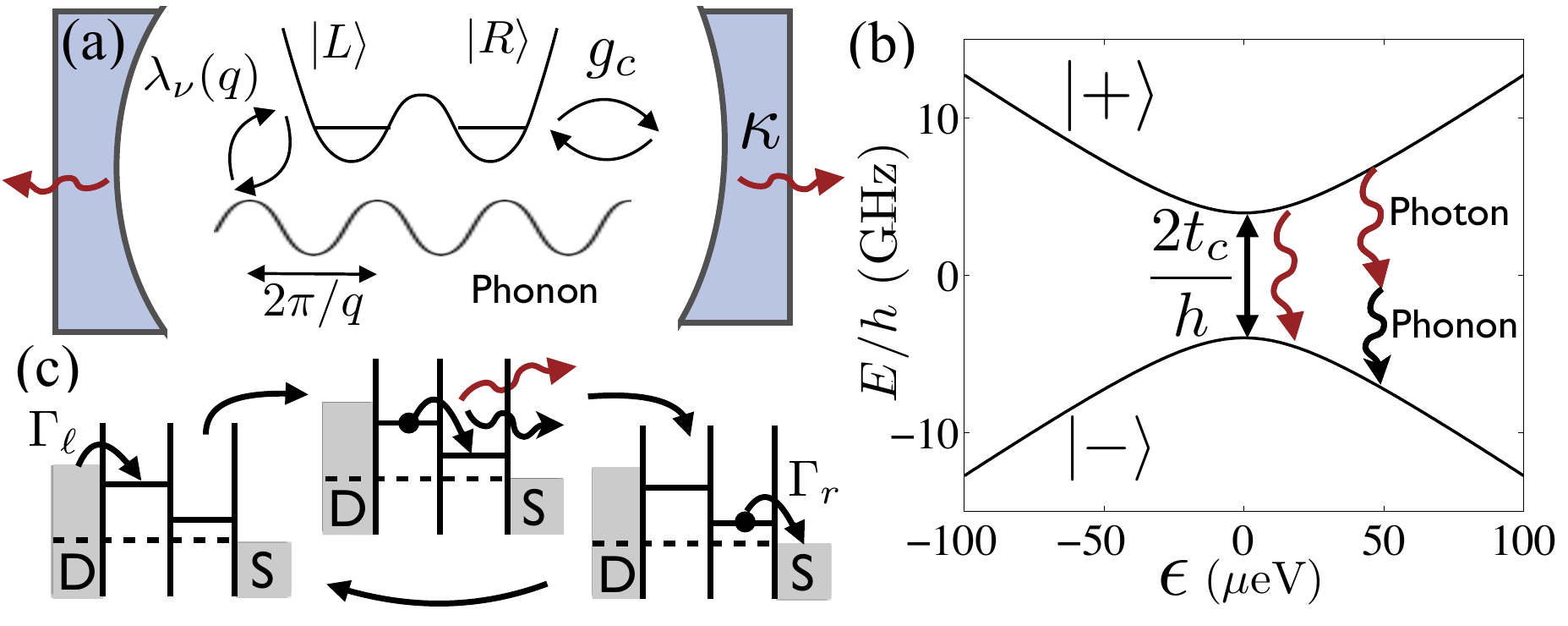}
\caption{(a) Schematic of combined DQD, cavity, and phonon system.    
(b) DQD energy spectrum versus detuning $\epsilon$ for $t_c =16.4~\mu$eV,  
  the direct photon emission process, and the phonon sideband for $\omega_c/2\pi = 8$~GHz.   
 (c) Transport cycle for DQD maser: a finite source drain bias leads to current flow via single electron tunneling. The interdot charge transition is accompanied by direct photon emission into the cavity (near zero detuning) 
 or a second order process involving emission of a photon and a phonon (large detuning). Electrons tunnel onto(off of) the left(right) dots at rate $\Gamma_{\ell(r)}$.
 } 
\label{default}
\end{center}
\end{figure}

\emph{DQD Gain Medium --} To analyze the {masing} process we first need to characterize the DQD gain medium and its coupling to the microwave cavity.   Following previous theoretical work \cite{Childress04,Jin11,Kulkarni14}, we  develop a simplified microscopic model for the system, consisting of a DQD, single mode cavity, phonon bath, and leads.
From this model we can  extract the gain, which is determined by three distinct quantities:  the DQD  photon emission and absorption rates and the population inversion of the DQD.   We then extend this result to multiple dots in the sideband-dominated limit.

Due to Coulomb blockade, each DQD can be restricted to two orbital states $\ket{L}$ and $\ket{R}$, where $\ket{L}$ has $(M+1,N)$ electrons and $\ket{R}$ has $(M,N+1)$ electrons in the (left, right) dots.  The different charge configurations of these states results in an electric dipole moment on the order of $ D\sim 1000 \, e\,  a_0$, where $e$ is the electronic charge and $a_0$ is the Bohr radius.  The Hamiltonian describing a single DQD coupled to a cavity is given by
 \begin{align} \label{eqn:H0}
 H_0 & =   \frac{\epsilon}{2} \sigma_z + t_c \sigma_x + \hbar \omega_c a^\dagger a + \hbar g_c \sigma_z (a+a^\dagger), 
\end{align}
where $\sigma_\mu$ are Pauli-matrices operating in the orbital subspace $\ket{L}$ and $\ket{R}$,$\epsilon$ is the detuning between the two dots, $t_c/\hbar$ is the {interdot tunneling rate}, $\omega_c$ is the cavity frequency, $g_c$ is the DQD-cavity coupling, and  $a^\dagger(a)$ are cavity photon creation(annihilation) operators, . 
The electron-phonon interaction takes the generic form 
$
H_{ep}/\hbar=  \sum_{q,\nu}   \omega_\nu(q)a_{q\nu}^\dagger a_{q\nu}+ \lambda_\nu(q) \sigma_z (a_{q\nu}^\dagger+a_{q\nu}),
$ where 
 $\omega_\nu(q)$ is the phonon dispersion, $\lambda_\nu(q)$ is a coupling constant that depends on momentum $q$ and mode index $\nu$ and $a_{q\nu}^\dagger(a_{q\nu})$ are phonon  creation(annihilation) operators.  The exact form of $\lambda_\nu(q)$ is set by the electronic wavefunctions, material properties, and boundary conditions. We focus on the phonon properties of nanowire QDs   \cite{Weber10}.

Diagonalizing the first two terms in $H_0$ leads to the eigenstates $\ket{\pm}$ 
\begin{align}
\ket{+} = \cos (\theta/2) \ket{L}-  \sin (\theta/2) \ket{R}, \\
\ket{-}= \sin (\theta/2) \ket{L}+  \cos (\theta/2) \ket{R}, 
\end{align}
where $\theta = \tan^{-1}(2t_c/\epsilon)$.   These states have an energy splitting $\hbar \omega_d= \sqrt{\epsilon^2+4\, t_c^2}$ shown in Fig.\ 1(b).    
Writing the Pauli-matrices in this new basis, the interaction between the DQD, phonons, and cavity photons is
\beu \label{eqn:H} 
H_{int} = \hbar (\cos \theta \sigma_z + \sin \theta \sigma_x) \big[ g_c\, a+ \sum_{q,\nu} \lambda_\nu(q) a_{q\nu} + h.c. \big].
\eeu
From this interaction we see that  both phonons and photons will cause relaxation
 from $\ket{+}$ to $\ket{-}$; 
 therefore, single electron tunneling through the dots will be correlated with photon and phonon emission \cite{Fujisawa98,Weber10}.  

In the presence of a finite source-drain bias, an electron first tunnels from the drain to the left dot, followed by an interdot charge transition from $\ket{L}$ to $\ket{R}$, and then leaves the right dot by tunneling to the source.
In the context of the maser, this can lead to population inversion when $\epsilon>0$ as it continually repumps $\ket{+}$.  In the limit where only single electrons can tunnel through the DQD, this process can be modeled by including a third, empty dot state $\ket{0}$ with incoherent tunneling rates $\Gamma_{\ell}$ from $\ket{0}\to \ket{L}$ and $\Gamma_r$ from $\ket{R} \to \ket{0}$ [see Fig.\ 1(c)].  Thus
the dynamics for a single DQD can be described by the master equation for the density matrix $\rho$  \cite{Jin11}
\begin{align*} \label{eqn:master}
\dot{\rho} &= - \frac{i}{\hbar} [H,\rho] + \kappa\, \mathcal{D}[a]\rho + \Gamma_\ell\, \mathcal{D}[\ket{L}\bra{0}] \rho  + \Gamma_r\, \mathcal{D}[\ket{0}\bra{R}]  \rho,
\end{align*} 
where $H=H_0+H_{ep}$ describes the coherent dynamics (including the phonons) and the incoherent evolution is described by the Lindblad super-operators $\mathcal{D}[A]\rho=-1/2 \{ A^\dagger A,\rho \} + A \rho A^\dagger$, for any operator $A$, corresponding to cavity decay, at rate $\kappa$, and inelastic electron tunneling.

Neglecting the phonons, the emission rate of photons into the cavity can be found perturbatively for small $g_c$ by using the Heisenberg-Langevin equations for the DQD-cavity system in a rotating wave approximation:
\begin{align}
\dot{\sigma}_- & = -(\Gamma+i (\omega_d-\omega_c))\, {\sigma_-} + i g_c\sin \theta\,  a\, \sigma_z + \sigma_z \mathcal{F}_d,\\
\dot{a} &=- \kappa/2\,  {a} +i g_c \sin \theta\, \sigma_- + \mathcal{F}_c.  
\end{align}
Here  $\mathcal{F}_{c(d)}$ are the associated noise operators for the cavity(dot) baths and $\Gamma$ is the total dephasing rate (defined below). 
\delete{due to the phonon bath and coupling to the leads,}      Adiabatic elimination and mean field theory, i.e., $\mean{a \sigma_z}\approx \mean{a}\mean{ \sigma_z}$, appropriate for large $\Gamma$, gives  the equation of motion for the cavity photon number $n_c= <a^\dagger a>$,
$
\dot{n}_c = - (\kappa - {R} \mean{\sigma_z} ) \langle {a^\dagger a} \rangle.
$
Where the direct photon emission rate for the DQD is
\be \label{eqn:game}
{R} \approx \frac{8 t_c^2}{\omega_d^2} \frac{g_c^2}{\Gamma^2 +  (\omega_d - \omega_c)^2} \Gamma.
\ee

The dominant effect of  phonons is to induce  relaxation from $\ket{+}$ to $\ket{-}$ via phonon emission.  Neglecting cavity effects, the zero-temperature emission rate is given by Fermi's Golden rule as
\be \label{eqn:gamd}
\gamma_d=  \frac{8 \pi\, t_c^2}{\omega_d^2} J(\omega_d) ,
\ee
where $J(\omega)= \sum_{q,\nu} \abs{\lambda_\nu(q)}^2 \delta(\omega_\nu(q)-\omega)$ is the spectral density of the phonons.  In the presence of thermal phonons with distribution $n_p(\omega)$, the total emission rate is $\gamma_d^\downarrow \equiv \gamma_d\, (n_p(\omega_d)+1)$. Starting in $\ket{-}$, there is also absorption at the rate $\gamma_d^\uparrow = \gamma_d\,  n_p(\omega_d)$. Using these expressions, we can write the total dephasing rate as $\Gamma = (\gamma_d^\downarrow+\gamma_d^\uparrow+ \Gamma_r)/2$.    We treat the phonon spectral density $J(\omega)$ using the microscopic model and measurements in Refs.\ \cite{Brandes05,Weber10}.  We take a 25 nm radius nanowire with a  separation between the two dots $d= 120$~nm, an axial confinement $a= 25$~nm for each dot, and a phonon speed of sound $c_n=4000$~m/s \cite{supp}.  


The effect of phonons on the photon emission is calculated by first performing a polaron transformation $H' = U H U^\dagger$ with \cite{Brandes05}
\be
U = e^{ \big[ g_c  (a- a^\dagger)/\omega_c + \sum_{q,\nu} \lambda_\nu(q) (a_{q\nu}-a_{q\nu}^\dagger)/\omega_\nu(q) \big] \cos \theta \sigma_z },
\ee
which removes the $\sigma_z$ terms in the interaction. The polaron transformation serves to dress the electronic states of the DQD with the ambient phonons in the environment. Perturbatively in $g_c/\omega_c$ and $\lambda_\nu(q)/\omega_\nu(q)$ this results in explicit terms in the Hamiltonian, which have not been considered in previous theoretical treatments of the DQD maser \cite{Childress04,Jin11,Kulkarni14}, describing second order photon-phonon processes 
\begin{align} \nonumber  
 H'&=\frac{4 t_c \epsilon}{\omega_d^2}  \sum_{q,\nu} \frac{i\, g_c \, \lambda_\nu(q)}{\omega_c\, \omega_\nu(q)} \Big[ (\omega_\nu(q)+\omega_c) (a \, a_{q\nu} - a^\dagger a_{q\nu}^\dagger) \\ \label{eqn:Hpea} 
 &+ (\omega_\nu(q)-\omega_c)(a \, a_{q\nu}^\dagger - a^\dagger a_{q\nu}) \Big] \sigma_y  +H'',
 \end{align}
where $H''$  contains terms that do not directly couple photons and phonons.  
The first term in Eq.\ (\ref{eqn:Hpea}) leads to phonon assisted emission [Fig.\ 1(b)], whereby {relaxation from $\ket{+}$ to $\ket{-}$ occurs by emitting} a phonon of frequency $\omega_\nu(q)=\omega_d-\omega_c$ along with a cavity photon.
{The second term leads to phonon assisted absorption, whereby relaxation occurs by emitting a phonon of frequency $\omega_\nu(q)=\omega_d+\omega_c$ and absorbing a cavity photon.}
Using Fermi's golden rule, these two terms give {the zero-temperature}, phonon-assisted-photon-emission $\gamma_e$ {and absorption $\gamma_a$} rates as 
\begin{align} \label{eqn:gampe} 
\gamma_{e} & \approx \frac{ 32\, \pi\, g_c^2\, \epsilon^2\,  t_c^2}{\omega_d^2\, \omega_c^2\, (\omega_d- \omega_c)^2} J(\omega_d-\omega_c), \\ \label{eqn:gampa}
\gamma_{a} & \approx \frac{ 32\, \pi\, g_c^2\, \epsilon^2\,  t_c^2}{\omega_d^2\, \omega_c^2\, (\omega_d+ \omega_c)^2} J(\omega_d+\omega_c).
\delete{\gamma_{e} & \approx \frac{ 32 \pi \epsilon^2\, g_c^2\, t_c^2}{\omega_d^2\, \omega_c^2 (\omega_d- \omega_c)^2} \gamma(\omega_d-\omega_c), \\ \label{eqn:gampa}}
\delete{\gamma_{a} &\approx \frac{ 4 \epsilon^2 g_c^2}{\omega_c^2 (\omega_d+\omega_c)^2} \gamma_{}(\omega_d + \omega_c).}
\end{align}
In the presence of thermal phonons, we define $\gamma_{e,a}^{\uparrow,\downarrow}$ analogously to case for the direct phonon process, {where $\downarrow$ refers to transitions from $\ket{+}$ to $\ket{-}$ and vice-versa for $\uparrow$.}  These thermal contributions are important because,
in addition to the ambient thermal phonons in the DQD, pumping current through the dot will  generate a large  population of phonons through Ohmic heating of the nanowire.  Since the equilibration time of the phonons is on the order of  picoseconds  ($\sim a/c_n \sim 10$~ps) and the cavity dynamics occur over a timescale of hundreds of nanoseconds ($\sim \kappa^{-1} \sim 100$ ns), we can  take the phonon bath to be in equilibrium with an effective temperature $T_{\textrm{eff}}$, such that $n_p(\omega) = (e^{\hbar \omega/k_B T_{\textrm{eff}}} -1)^{-1}$.  In Ref.~\cite{Liu14a}, $T_\textrm{eff}$ was estimated to be as high as 1~K due to the large nA currents flowing through the nanowire.

 \begin{figure}[t]
\includegraphics[width=.47 \textwidth]{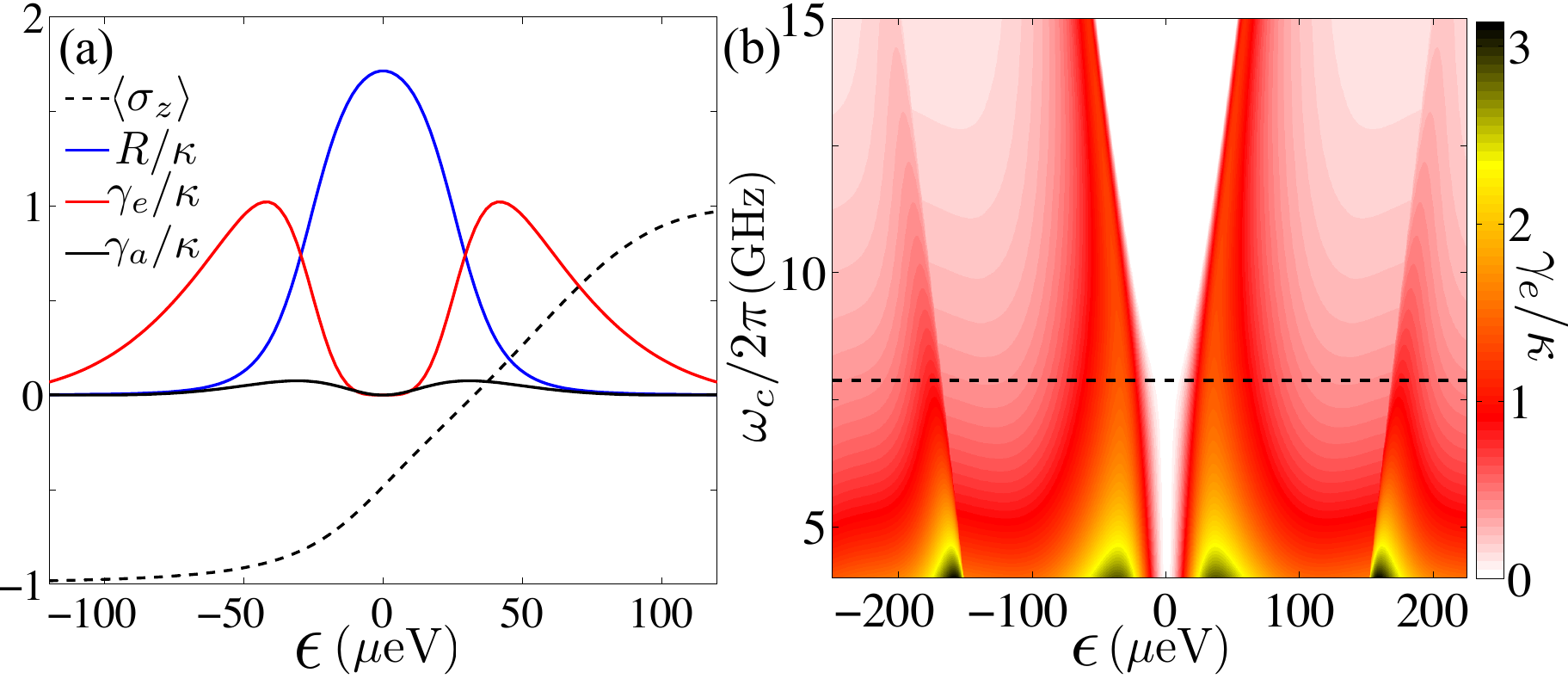}
\caption{ 
(a) Population inversion $\mean{\sigma_z}$, direct photon rate  $R$, and  phonon-assisted emission $\gamma_{e}$ and absorption $\gamma_{a}$ rates as a function of $\epsilon$ with $t_c=16.4~\mu$eV, $g_c/2\pi = 90$~MHz, $\omega_c/2\pi = 8$~GHz, $J(2 t_c/\hbar)=5$~GHz, $\Gamma_\ell/2\pi = \Gamma_r/2\pi =  4$~GHz, and $T_{\textrm{eff}}=0$.  
 (b) $\gamma_e/ \kappa$ plotted as a function $\epsilon$ and $\omega_c/2\pi$.
 The dashed line corresponds to $\omega_c/2\pi = 8$~GHz used in Ref.\ \cite{Liu14b}. 
}
\label{default}
\end{figure}

Figure 2(a) shows the key quantities in determining the gain for parameters similar to Ref.\ \cite{Liu14a,Liu14b}: the population inversion $\mean{\sigma_z}$, obtained from Eq.\ (5) in the absence of the cavity, and the various photon emission and absorption rates. For $\epsilon=0$ the DQD eigenstates are equal admixtures of $\ket{L}$ and $\ket{R}$ and the inversion is small, while for large, positive $\epsilon$,  $\ket{+} \to \ket{L}$, the system becomes completely inverted as seen in Fig.\ 2(a). Although $R$ and $\gamma_e$ are comparable in magnitude, $R$ dominates at small detunings, where the population inversion is small, and $\gamma_e$ dominates at large detunings, where there is large population inversion.  Based on Eq.\ (\ref{eqn:game}-\ref{eqn:gampe}), it is possible for $R$ to dominate the gain at large $\epsilon$ when $\omega_c \gg 2 t_c$.  However, this analysis has so far neglected charge noise in the system.  In Ref.\ \cite{Petersson12,Liu14a}, this was estimated to lead to slowly varying noise in $\epsilon$ with an rms value of $ (20-40)~\mu$eV~$\gg \hbar \Gamma \approx 1~\mu$eV.  In the presence of such large noise, $\gamma_e$ will dominate over $R$ for large $\epsilon$. 
Figure 2(b) shows $\gamma_e$ for varying $\epsilon$ and $\omega_c$.  Because the phonon-assisted process is perturbative in $g_c/\omega_c$, it has the strongest effect for small cavity frequencies.   The second peak at $\epsilon =150~\mu$eV arises from the second  phonon branch in the nanowire.


\emph{DQD Maser --}
Away from the masing threshold, we can find the response of the system within mean field theory. 
Including thermal effects, the Heisenberg-Langevin equations gives rise to the mean field equations for the field amplitude $\alpha = \mean{a}$ and the population in the upper state $u= \mean{\ket{+}\bra{+}}$ \cite{MeystreBook,supp}:
\begin{align} \label{eqn:seom1}
\dot{\alpha} &= - \big( [\kappa - g(u) ] /2 + i \delta\big) \alpha + \Omega, \\ \label{eqn:seom2}
\dot{u} &= \Gamma_p (u_0- u) - S(u) \abs{\alpha}^2,
 \end{align}
 where we have defined the gain rate function $g(u)$ and a saturation function $S(u)$ as 
 \begin{align} \label{eq:gu0}
g&= {R}\, (2 u-1) + (\gamma^\downarrow_{e}-\gamma^\downarrow_{a}) u - (\gamma_{e}^\uparrow - \gamma_{a}^\uparrow) (1-u), \\
S &={R}\, (2 u-1) + (\gamma^\downarrow_{e}+\gamma^\downarrow_{a}) u - (\gamma_{e}^\uparrow + \gamma_{a}^\uparrow) (1-u).
\end{align}
Here we have introduced the drive with amplitude $\Omega$ and frequency $\omega_\ell$.  The detuning 
$\delta= \omega_c - \omega_\ell - R \Gamma (2 u-1)/2\Delta$, includes the cavity line pulling \cite{LaserBook}.  
 $\Gamma_p$ and $u_0$ are the effective pumping rate and upper state population, respectively.
 The full expressions are given in Ref.\ \cite{Childress04,Jin11,Kulkarni14}. 
For large $\epsilon$ and $\Gamma_{\ell,r}$, they reduce to  $\Gamma_p \approx \Gamma_\ell \Gamma_r/(\Gamma_\ell+2 \Gamma_r)$ and $u_0\approx 1 - 2(\gamma_d^\downarrow+\gamma_d^\uparrow)/\Gamma_p$.
In the case of the experiment, where there are multiple DQD (two), the large dephasing rate $\Gamma$ allows  $u$ to simply be replaced by the average upper state population in each DQD and  $g$ to be multiplied by the number of DQDs.

 \begin{figure}[t]
\begin{center}
\includegraphics[width=.49 \textwidth]{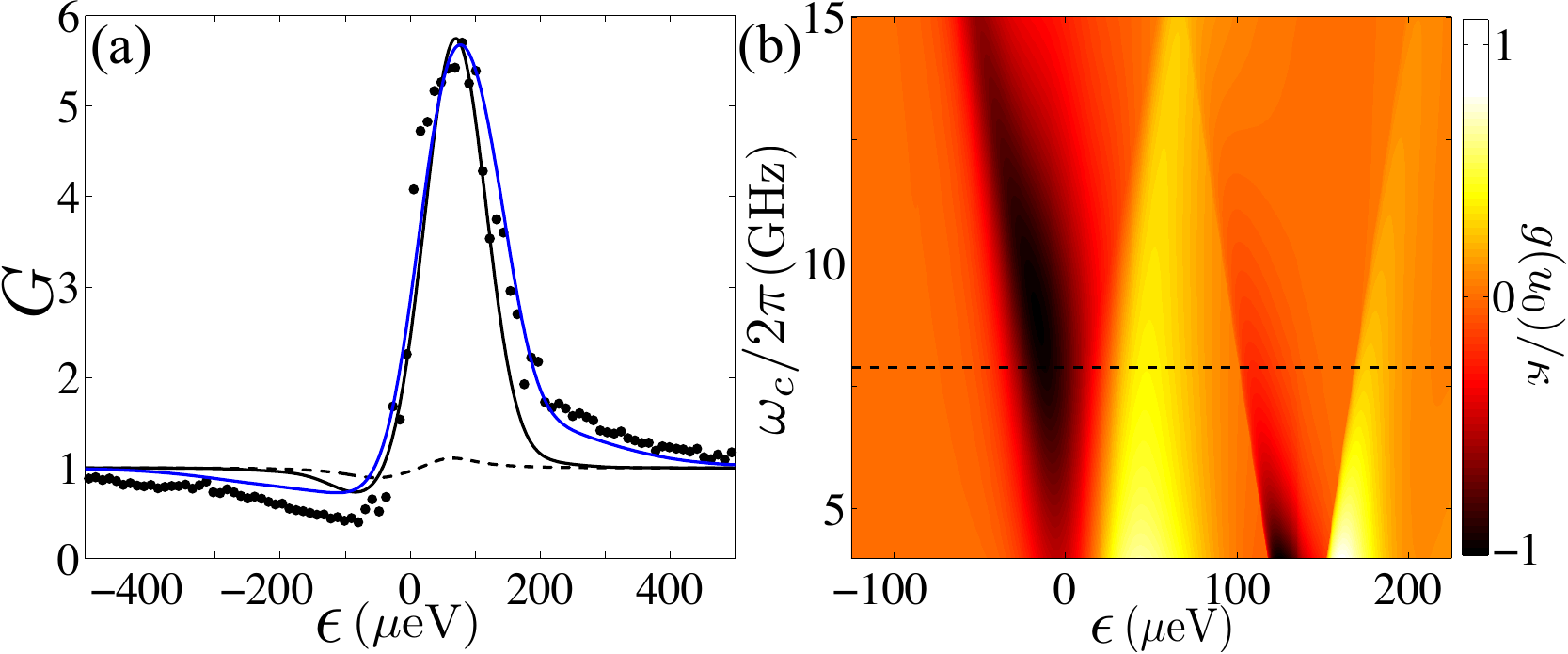}
\caption{
  (a) (Circles) Experimentally measured gain in one of the DQDs with  $\omega_c/2\pi = 8$~GHz, $\kappa/2\pi=2.6$~MHz, $t_c = 50(10)~\mu$eV, $\Gamma_\ell/2\pi = \Gamma_r/2\pi =  17(2)$~GHz, $T_{\textrm{eff}}=3(1)$~K, and $g_c/2\pi = 100(20)$~MHz.  (Black) Fit to the theory with $J(2 t_c/\hbar) = 2.4(2)$ GHz a free parameter and only including contributions from the lowest phonon branch.  (Blue)  Includes contributions to $J(\omega)$ from the second phonon mode and substrate phonons \cite{supp}.  (Dashed) Gain neglecting the phonon-assisted contributions.  (b) Gain rate function $g(u_0)$ plotted versus $\epsilon$ and $\omega_c/2\pi$ with parameters as in Fig.\ 2(a).  The second peak near $\epsilon =150~\mu$eV arises from the second phonon mode in the nanowire.
 }
\label{default}
\end{center}
\end{figure}

For weak driving fields and below threshold operation, the normalized gain $\abs{\alpha(\delta;g)}^2/\abs{\alpha(0;0)}^2$ is given by 
\be \label{eqn:G}
G (\delta) = \frac{\kappa^2}{\big[ \kappa - g(u_0) \big]^2 + 4 \delta^2}.
\ee
From transmission measurements it is known $\omega_c/2\pi = 8$~GHz and $\kappa/2\pi=2.6$~MHz \cite{Liu14b},   modeling the current through the dot at finite bias gives $t_c = 50(10)~\mu$eV, $\Gamma_\ell/2\pi = \Gamma_r/2\pi =  17(2)$~GHz, $T_{\textrm{eff}}=3(1)$~K \cite{Liu14a}, and the gain at zero bias gives $g_c/2\pi=100(20)$~MHz \cite{Liu14a,footnote}.  
To account for charge noise, we convolved the gain with a Gaussian of width $40(10)~\mu$eV \cite{Petersson12}.  Finally, we find $J(2 t_c/\hbar)=2.4(2)$~GHz by fitting the gain at finite bias  including only the first order phonon branch in $J(\omega)$ (which is a valid approximation for $\abs{\epsilon}<200~\mu$eV for a 25 nm radius nanowire \cite{Weber10}).  To match the broad tails in the gain data for $\epsilon > 200~\mu$eV [see Fig.\ 3(a)], we include the coupling to the second  longitudinal mode and substrate phonons \cite{supp}.

Figure 3(a) shows the comparison between the measured $G(0)$ for a single DQD  and a fit to our model.    From the data we can conclusively rule out a model with just the direct photon emission process as it would require a DQD-cavity coupling $g_c$ 10-100 times larger than what was measured.  
On the other hand,  when the  phonon-assisted processes are included, we find good agreement.

Equations (\ref{eqn:seom1})-(\ref{eqn:seom2}) predict a masing transition when $g(u_0) \gtrsim \kappa$.  This is consistent with the experimental results, where there are two DQDs in the cavity, each with peak gain rates slightly below the cavity linewidth.  When only one DQD is configured to maximum gain, no significant photon emission is observed;  however, when both are tuned to maximum gain, such that  the combined gain rate  is greater than the cavity linewidth, masing is observed \cite{Liu14b}.  

Because of the strong dependence of the gain on the phonon-assisted process, measuring the gain near threshold is a sensitive measurement of the phonon spectral density $J(\omega)$.  In particular, by tuning $\omega_c$ and measuring the gain curves as in Fig.\ 3(a), one could precisely determine the frequency dependence of $J(\omega)$ by extracting $g(u_0)$.  This is illustrated in Fig.\ 3(b), which shows $g(u_0)$ at $T_{\textrm{eff}}=0$ for varying $\epsilon$ and $\omega_c$, where we see that there will be a second  peak in the gain at low frequencies when $\omega_d - \omega_c$ equals the gap to the second longitudinal phonon mode of the nanowire \cite{supp}.


Finally, this work shows that phonons will be important for circuit quantum electrodynamics (QED) experiments involving spin-photon entanglement and the generation of non-classical states of light.
 In conventional cavity QED, the fidelity of these operations is  limited by the largeness of the Purcell factor $g_c^2/\kappa\, \Gamma$ \cite{Kimble08}.  In the case of the DQD, as $g_c$ approaches $\omega_c$ the phonon-assisted processes can dominate over the bare relaxation rate $\Gamma$.   This will ultimately constrain the fidelity of these operations, but it  also represents an unexplored regime of cavity QED that is unique to the solid-state environment and energy scales of the DQD system.


\begin{acknowledgements}
\emph{Acknowledgements --} We thank G. Solomon and V. Srinivasa for illuminating discussions. Research at Princeton was supported by the Packard Foundation and the National Science Foundation (DMR-1409556 and DMR-1420541), DARPA QuEST (HR0011-09-1-0007) and ARO (W911NF-08-1-0189).
\end{acknowledgements}

\bibliographystyle{../apsrev-nourl}
\bibliography{../DDMaser}

\begin{thebibliography}{33}
\expandafter\ifx\csname natexlab\endcsname\relax\def\natexlab#1{#1}\fi
\expandafter\ifx\csname bibnamefont\endcsname\relax
  \def\bibnamefont#1{#1}\fi
\expandafter\ifx\csname bibfnamefont\endcsname\relax
  \def\bibfnamefont#1{#1}\fi
\expandafter\ifx\csname citenamefont\endcsname\relax
  \def\citenamefont#1{#1}\fi
\expandafter\ifx\csname url\endcsname\relax
  \def\url#1{\texttt{#1}}\fi
\expandafter\ifx\csname urlprefix\endcsname\relax\def\urlprefix{URL }\fi
\providecommand{\bibinfo}[2]{#2}
\providecommand{\eprint}[2][]{\url{#2}}

\bibitem[{\citenamefont{Sargent et~al.}(1978)\citenamefont{Sargent, Scully, and
  Lamb}}]{LaserBook}
\bibinfo{author}{\bibfnamefont{M.}~\bibnamefont{Sargent}},
  \bibinfo{author}{\bibfnamefont{M.}~\bibnamefont{Scully}}, \bibnamefont{and}
  \bibinfo{author}{\bibfnamefont{W.}~\bibnamefont{Lamb}},
  \emph{\bibinfo{title}{Laser Physics}} (\bibinfo{publisher}{Perseus Books
  Group}, \bibinfo{year}{1978}).

\bibitem[{\citenamefont{Mu and Savage}(1992)}]{Mu92}
\bibinfo{author}{\bibfnamefont{Y.}~\bibnamefont{Mu}} \bibnamefont{and}
  \bibinfo{author}{\bibfnamefont{C.~M.} \bibnamefont{Savage}},
  \bibinfo{journal}{Phys. Rev. A} \textbf{\bibinfo{volume}{46}},
  \bibinfo{pages}{5944} (\bibinfo{year}{1992}).

\bibitem[{\citenamefont{Rice and Carmichael}(1994)}]{Rice94}
\bibinfo{author}{\bibfnamefont{P.~R.} \bibnamefont{Rice}} \bibnamefont{and}
  \bibinfo{author}{\bibfnamefont{H.~J.} \bibnamefont{Carmichael}},
  \bibinfo{journal}{Phys. Rev. A} \textbf{\bibinfo{volume}{50}},
  \bibinfo{pages}{4318} (\bibinfo{year}{1994}).

\bibitem[{\citenamefont{Bj{\"o}rk et~al.}(1994)\citenamefont{Bj{\"o}rk,
  Karlsson, and Yamamoto}}]{Bjork94}
\bibinfo{author}{\bibfnamefont{G.}~\bibnamefont{Bj{\"o}rk}},
  \bibinfo{author}{\bibfnamefont{A.}~\bibnamefont{Karlsson}}, \bibnamefont{and}
  \bibinfo{author}{\bibfnamefont{Y.}~\bibnamefont{Yamamoto}},
  \bibinfo{journal}{Phys. Rev. A} \textbf{\bibinfo{volume}{50}},
  \bibinfo{pages}{1675} (\bibinfo{year}{1994}).

\bibitem[{\citenamefont{McKeever et~al.}(2003)\citenamefont{McKeever, Boca,
  Boozer, Buck, and Kimble}}]{McKeever03}
\bibinfo{author}{\bibfnamefont{J.}~\bibnamefont{McKeever}},
  \bibinfo{author}{\bibfnamefont{A.}~\bibnamefont{Boca}},
  \bibinfo{author}{\bibfnamefont{A.~D.} \bibnamefont{Boozer}},
  \bibinfo{author}{\bibfnamefont{J.~R.} \bibnamefont{Buck}}, \bibnamefont{and}
  \bibinfo{author}{\bibfnamefont{H.~J.} \bibnamefont{Kimble}},
  \bibinfo{journal}{Nature (London)} \textbf{\bibinfo{volume}{425}},
  \bibinfo{pages}{268} (\bibinfo{year}{2003}).

\bibitem[{\citenamefont{Walther et~al.}(2006)\citenamefont{Walther, Varcoe,
  Englert, and Becker}}]{Walther06}
\bibinfo{author}{\bibfnamefont{H.}~\bibnamefont{Walther}},
  \bibinfo{author}{\bibfnamefont{B.~T.~H.} \bibnamefont{Varcoe}},
  \bibinfo{author}{\bibfnamefont{B.-G.} \bibnamefont{Englert}},
  \bibnamefont{and} \bibinfo{author}{\bibfnamefont{T.}~\bibnamefont{Becker}},
  \bibinfo{journal}{Rep. Prog. Phys.} \textbf{\bibinfo{volume}{69}},
  \bibinfo{pages}{1325} (\bibinfo{year}{2006}).

\bibitem[{\citenamefont{Xie et~al.}(2007)\citenamefont{Xie, Gotzinger, Fang,
  Cao, and Solomon}}]{Xie07}
\bibinfo{author}{\bibfnamefont{Z.~G.} \bibnamefont{Xie}},
  \bibinfo{author}{\bibfnamefont{S.}~\bibnamefont{Gotzinger}},
  \bibinfo{author}{\bibfnamefont{W.}~\bibnamefont{Fang}},
  \bibinfo{author}{\bibfnamefont{H.}~\bibnamefont{Cao}}, \bibnamefont{and}
  \bibinfo{author}{\bibfnamefont{G.~S.} \bibnamefont{Solomon}},
  \bibinfo{journal}{Phys. Rev. Lett.} \textbf{\bibinfo{volume}{98}},
  \bibinfo{pages}{117401} (\bibinfo{year}{2007}).

\bibitem[{\citenamefont{Nomura et~al.}(2010)\citenamefont{Nomura, Kumagai,
  Iwamoto, Ota, and Arakawa}}]{Nomura10}
\bibinfo{author}{\bibfnamefont{M.}~\bibnamefont{Nomura}},
  \bibinfo{author}{\bibfnamefont{N.}~\bibnamefont{Kumagai}},
  \bibinfo{author}{\bibfnamefont{S.}~\bibnamefont{Iwamoto}},
  \bibinfo{author}{\bibfnamefont{Y.}~\bibnamefont{Ota}}, \bibnamefont{and}
  \bibinfo{author}{\bibfnamefont{Y.}~\bibnamefont{Arakawa}},
  \bibinfo{journal}{Nat. Phys.} \textbf{\bibinfo{volume}{6}},
  \bibinfo{pages}{279} (\bibinfo{year}{2010}).

\bibitem[{\citenamefont{Astafiev et~al.}(2007)\citenamefont{Astafiev, Inomata,
  Niskanen, Yamamoto, Pashkin, Nakamura, and Tsai}}]{Nakamura07}
\bibinfo{author}{\bibfnamefont{O.}~\bibnamefont{Astafiev}},
  \bibinfo{author}{\bibfnamefont{K.}~\bibnamefont{Inomata}},
  \bibinfo{author}{\bibfnamefont{A.}~\bibnamefont{Niskanen}},
  \bibinfo{author}{\bibfnamefont{T.}~\bibnamefont{Yamamoto}},
  \bibinfo{author}{\bibfnamefont{Y.~A.} \bibnamefont{Pashkin}},
  \bibinfo{author}{\bibfnamefont{Y.}~\bibnamefont{Nakamura}}, \bibnamefont{and}
  \bibinfo{author}{\bibfnamefont{J.}~\bibnamefont{Tsai}},
  \bibinfo{journal}{Nature (London)} \textbf{\bibinfo{volume}{449}},
  \bibinfo{pages}{588} (\bibinfo{year}{2007}).

\bibitem[{\citenamefont{Chen et~al.}(2014)\citenamefont{Chen, Li, Armour,
  Brahimi, Stettenheim, Sirois, Simmonds, Blencowe, and Rimberg}}]{Chen14}
\bibinfo{author}{\bibfnamefont{F.}~\bibnamefont{Chen}},
  \bibinfo{author}{\bibfnamefont{J.}~\bibnamefont{Li}},
  \bibinfo{author}{\bibfnamefont{A.~D.} \bibnamefont{Armour}},
  \bibinfo{author}{\bibfnamefont{E.}~\bibnamefont{Brahimi}},
  \bibinfo{author}{\bibfnamefont{J.}~\bibnamefont{Stettenheim}},
  \bibinfo{author}{\bibfnamefont{A.~J.} \bibnamefont{Sirois}},
  \bibinfo{author}{\bibfnamefont{R.~W.} \bibnamefont{Simmonds}},
  \bibinfo{author}{\bibfnamefont{M.~P.} \bibnamefont{Blencowe}},
  \bibnamefont{and} \bibinfo{author}{\bibfnamefont{A.~J.}
  \bibnamefont{Rimberg}}, \bibinfo{journal}{Phys. Rev. B}
  \textbf{\bibinfo{volume}{90}}, \bibinfo{pages}{020506}
  (\bibinfo{year}{2014}).

\bibitem[{\citenamefont{Childress et~al.}(2004)\citenamefont{Childress,
  Sorensen, and Lukin}}]{Childress04}
\bibinfo{author}{\bibfnamefont{L.}~\bibnamefont{Childress}},
  \bibinfo{author}{\bibfnamefont{A.~S.} \bibnamefont{Sorensen}},
  \bibnamefont{and} \bibinfo{author}{\bibfnamefont{M.~D.} \bibnamefont{Lukin}},
  \bibinfo{journal}{Phys. Rev. A} \textbf{\bibinfo{volume}{69}},
  \bibinfo{pages}{042302} (\bibinfo{year}{2004}).

\bibitem[{\citenamefont{Jin et~al.}(2011)\citenamefont{Jin, Marthaler, Cole,
  Shnirman, and Sch{\"o}n}}]{Jin11}
\bibinfo{author}{\bibfnamefont{P.-Q.} \bibnamefont{Jin}},
  \bibinfo{author}{\bibfnamefont{M.}~\bibnamefont{Marthaler}},
  \bibinfo{author}{\bibfnamefont{J.~H.} \bibnamefont{Cole}},
  \bibinfo{author}{\bibfnamefont{A.}~\bibnamefont{Shnirman}}, \bibnamefont{and}
  \bibinfo{author}{\bibfnamefont{G.}~\bibnamefont{Sch{\"o}n}},
  \bibinfo{journal}{Phys. Rev. B} \textbf{\bibinfo{volume}{84}},
  \bibinfo{pages}{035322} (\bibinfo{year}{2011}).

\bibitem[{\citenamefont{Kulkarni et~al.}(2014)\citenamefont{Kulkarni, Cotlet,
  and T{\"u}reci}}]{Kulkarni14}
\bibinfo{author}{\bibfnamefont{M.}~\bibnamefont{Kulkarni}},
  \bibinfo{author}{\bibfnamefont{O.}~\bibnamefont{Cotlet}}, \bibnamefont{and}
  \bibinfo{author}{\bibfnamefont{H.~E.} \bibnamefont{T{\"u}reci}},
  \bibinfo{journal}{Phys. Rev. B} \textbf{\bibinfo{volume}{90}},
  \bibinfo{pages}{125402} (\bibinfo{year}{2014}).

\bibitem[{\citenamefont{Liu et~al.}(2014)\citenamefont{Liu, Petersson, Stehlik,
  Taylor, and Petta}}]{Liu14a}
\bibinfo{author}{\bibfnamefont{Y.-Y.} \bibnamefont{Liu}},
  \bibinfo{author}{\bibfnamefont{K.~D.} \bibnamefont{Petersson}},
  \bibinfo{author}{\bibfnamefont{J.}~\bibnamefont{Stehlik}},
  \bibinfo{author}{\bibfnamefont{J.~M.} \bibnamefont{Taylor}},
  \bibnamefont{and} \bibinfo{author}{\bibfnamefont{J.~R.} \bibnamefont{Petta}},
  \bibinfo{journal}{Phys. Rev. Lett.} \textbf{\bibinfo{volume}{113}},
  \bibinfo{pages}{036801} (\bibinfo{year}{2014}).

\bibitem[{\citenamefont{Liu et~al.}(2015)\citenamefont{Liu, Stehlik, Eichler,
  Gullans, Taylor, and Petta}}]{Liu14b}
\bibinfo{author}{\bibfnamefont{Y.~Y.} \bibnamefont{Liu}},
  \bibinfo{author}{\bibfnamefont{J.}~\bibnamefont{Stehlik}},
  \bibinfo{author}{\bibfnamefont{C.}~\bibnamefont{Eichler}},
  \bibinfo{author}{\bibfnamefont{M.~J.} \bibnamefont{Gullans}},
  \bibinfo{author}{\bibfnamefont{J.~M.} \bibnamefont{Taylor}},
  \bibnamefont{and} \bibinfo{author}{\bibfnamefont{J.~R.} \bibnamefont{Petta}},
  \bibinfo{journal}{Science} \textbf{\bibinfo{volume}{347}},
  \bibinfo{pages}{285} (\bibinfo{year}{2015}).

\bibitem[{\citenamefont{Meystre and Sargent}(2007)}]{MeystreBook}
\bibinfo{author}{\bibfnamefont{P.}~\bibnamefont{Meystre}} \bibnamefont{and}
  \bibinfo{author}{\bibfnamefont{M.}~\bibnamefont{Sargent}},
  \emph{\bibinfo{title}{Elements of quantum optics}}, vol.~\bibinfo{volume}{3}
  (\bibinfo{publisher}{Springer Berlin}, \bibinfo{year}{2007}).

\bibitem[{\citenamefont{Kimble}(2008)}]{Kimble08}
\bibinfo{author}{\bibfnamefont{H.~J.} \bibnamefont{Kimble}},
  \bibinfo{journal}{Nature} \textbf{\bibinfo{volume}{453}},
  \bibinfo{pages}{1023} (\bibinfo{year}{2008}).

\bibitem[{\citenamefont{Petersson et~al.}(2012)\citenamefont{Petersson, McFaul,
  Schroer, Jung, Taylor, Houck, and Petta}}]{Petersson12}
\bibinfo{author}{\bibfnamefont{K.~D.} \bibnamefont{Petersson}},
  \bibinfo{author}{\bibfnamefont{L.~W.} \bibnamefont{McFaul}},
  \bibinfo{author}{\bibfnamefont{M.~D.} \bibnamefont{Schroer}},
  \bibinfo{author}{\bibfnamefont{M.}~\bibnamefont{Jung}},
  \bibinfo{author}{\bibfnamefont{J.~M.} \bibnamefont{Taylor}},
  \bibinfo{author}{\bibfnamefont{A.~A.} \bibnamefont{Houck}}, \bibnamefont{and}
  \bibinfo{author}{\bibfnamefont{J.~R.} \bibnamefont{Petta}},
  \bibinfo{journal}{Nature (London)} \textbf{\bibinfo{volume}{490}},
  \bibinfo{pages}{380} (\bibinfo{year}{2012}).

\bibitem[{\citenamefont{Basset et~al.}(2013)\citenamefont{Basset, Jarausch,
  Stockklauser, Frey, Reichl, Wegscheider, Ihn, Ensslin, and
  Wallraff}}]{Bassett13}
\bibinfo{author}{\bibfnamefont{J.}~\bibnamefont{Basset}},
  \bibinfo{author}{\bibfnamefont{D.~D.} \bibnamefont{Jarausch}},
  \bibinfo{author}{\bibfnamefont{A.}~\bibnamefont{Stockklauser}},
  \bibinfo{author}{\bibfnamefont{T.}~\bibnamefont{Frey}},
  \bibinfo{author}{\bibfnamefont{C.}~\bibnamefont{Reichl}},
  \bibinfo{author}{\bibfnamefont{W.}~\bibnamefont{Wegscheider}},
  \bibinfo{author}{\bibfnamefont{T.~M.} \bibnamefont{Ihn}},
  \bibinfo{author}{\bibfnamefont{K.}~\bibnamefont{Ensslin}}, \bibnamefont{and}
  \bibinfo{author}{\bibfnamefont{A.}~\bibnamefont{Wallraff}},
  \bibinfo{journal}{Phys. Rev. B} \textbf{\bibinfo{volume}{88}},
  \bibinfo{pages}{125312} (\bibinfo{year}{2013}).

\bibitem[{Den()}]{Deng14b}
\bibinfo{note}{G.-W. Deng, D. Wa, S.-X. Li, J. R. Johansson, W.-C. Kong, H.-O.
  Li, G. Cao, M. Xiao, G.-C. Guo, F. Nori, H.-W. Jiang, and G.-P. Guo,
  arXiv:1409.4980.}

\bibitem[{\citenamefont{Bergenfeldt and Samuelsson}(2013)}]{Bergenfeldt13}
\bibinfo{author}{\bibfnamefont{C.}~\bibnamefont{Bergenfeldt}} \bibnamefont{and}
  \bibinfo{author}{\bibfnamefont{P.}~\bibnamefont{Samuelsson}},
  \bibinfo{journal}{Phys. Rev. B} \textbf{\bibinfo{volume}{87}},
  \bibinfo{pages}{195427} (\bibinfo{year}{2013}).

\bibitem[{\citenamefont{Nielsen et~al.}(2004)\citenamefont{Nielsen, Gartner,
  and Jahnke}}]{Nielsen04}
\bibinfo{author}{\bibfnamefont{T.~R.} \bibnamefont{Nielsen}},
  \bibinfo{author}{\bibfnamefont{P.}~\bibnamefont{Gartner}}, \bibnamefont{and}
  \bibinfo{author}{\bibfnamefont{F.}~\bibnamefont{Jahnke}},
  \bibinfo{journal}{Phys. Rev. B} \textbf{\bibinfo{volume}{69}},
  \bibinfo{pages}{235314} (\bibinfo{year}{2004}).

\bibitem[{\citenamefont{Wijnen et~al.}(2008)\citenamefont{Wijnen, Blokland,
  Chin, Christianen, and Maan}}]{Wijnen08}
\bibinfo{author}{\bibfnamefont{F.~J.~P.} \bibnamefont{Wijnen}},
  \bibinfo{author}{\bibfnamefont{J.~H.} \bibnamefont{Blokland}},
  \bibinfo{author}{\bibfnamefont{P.~T.~K.} \bibnamefont{Chin}},
  \bibinfo{author}{\bibfnamefont{P.~C.~M.} \bibnamefont{Christianen}},
  \bibnamefont{and} \bibinfo{author}{\bibfnamefont{J.~C.} \bibnamefont{Maan}},
  \bibinfo{journal}{Phys. Rev. B} \textbf{\bibinfo{volume}{78}},
  \bibinfo{pages}{235318} (\bibinfo{year}{2008}).

\bibitem[{\citenamefont{Majumdar et~al.}(2011)\citenamefont{Majumdar, Kim,
  Gong, Bajcsy, and Vu\ifmmode \check{c}\else
  \v{c}\fi{}kovi\ifmmode~\acute{c}\else \'{c}\fi{}}}]{Majumdar11}
\bibinfo{author}{\bibfnamefont{A.}~\bibnamefont{Majumdar}},
  \bibinfo{author}{\bibfnamefont{E.~D.} \bibnamefont{Kim}},
  \bibinfo{author}{\bibfnamefont{Y.}~\bibnamefont{Gong}},
  \bibinfo{author}{\bibfnamefont{M.}~\bibnamefont{Bajcsy}}, \bibnamefont{and}
  \bibinfo{author}{\bibfnamefont{J.}~\bibnamefont{Vu\ifmmode \check{c}\else
  \v{c}\fi{}kovi\ifmmode~\acute{c}\else \'{c}\fi{}}}, \bibinfo{journal}{Phys.
  Rev. B} \textbf{\bibinfo{volume}{84}}, \bibinfo{pages}{085309}
  (\bibinfo{year}{2011}).

\bibitem[{\citenamefont{Quilter et~al.}(2015)\citenamefont{Quilter, Brash, Liu,
  Gl{\"a}ssl, Barth, Axt, Ramsay, Skolnick, and Fox}}]{Quilter15}
\bibinfo{author}{\bibfnamefont{J.~H.} \bibnamefont{Quilter}},
  \bibinfo{author}{\bibfnamefont{A.~J.} \bibnamefont{Brash}},
  \bibinfo{author}{\bibfnamefont{F.}~\bibnamefont{Liu}},
  \bibinfo{author}{\bibfnamefont{M.}~\bibnamefont{Gl{\"a}ssl}},
  \bibinfo{author}{\bibfnamefont{A.~M.} \bibnamefont{Barth}},
  \bibinfo{author}{\bibfnamefont{V.~M.} \bibnamefont{Axt}},
  \bibinfo{author}{\bibfnamefont{A.~J.} \bibnamefont{Ramsay}},
  \bibinfo{author}{\bibfnamefont{M.~S.} \bibnamefont{Skolnick}},
  \bibnamefont{and} \bibinfo{author}{\bibfnamefont{A.~M.} \bibnamefont{Fox}},
  \bibinfo{journal}{Phys. Rev. Lett.} \textbf{\bibinfo{volume}{114}},
  \bibinfo{pages}{137401} (\bibinfo{year}{2015}).

\bibitem[{\citenamefont{Weber et~al.}(2010)\citenamefont{Weber, Fuhrer, Fasth,
  Lindwall, Samuelson, and Wacker}}]{Weber10}
\bibinfo{author}{\bibfnamefont{C.}~\bibnamefont{Weber}},
  \bibinfo{author}{\bibfnamefont{A.}~\bibnamefont{Fuhrer}},
  \bibinfo{author}{\bibfnamefont{C.}~\bibnamefont{Fasth}},
  \bibinfo{author}{\bibfnamefont{G.}~\bibnamefont{Lindwall}},
  \bibinfo{author}{\bibfnamefont{L.}~\bibnamefont{Samuelson}},
  \bibnamefont{and} \bibinfo{author}{\bibfnamefont{A.}~\bibnamefont{Wacker}},
  \bibinfo{journal}{Phys. Rev. Lett.} \textbf{\bibinfo{volume}{104}},
  \bibinfo{pages}{036801} (\bibinfo{year}{2010}).

\bibitem[{\citenamefont{Fujisawa et~al.}(1998)\citenamefont{Fujisawa,
  Oosterkamp, van~der Wiel, Broer, Aguado, Tarucha, and
  Kouwenhoven}}]{Fujisawa98}
\bibinfo{author}{\bibfnamefont{T.}~\bibnamefont{Fujisawa}},
  \bibinfo{author}{\bibfnamefont{T.~H.} \bibnamefont{Oosterkamp}},
  \bibinfo{author}{\bibfnamefont{W.~G.} \bibnamefont{van~der Wiel}},
  \bibinfo{author}{\bibfnamefont{B.~W.} \bibnamefont{Broer}},
  \bibinfo{author}{\bibfnamefont{R.}~\bibnamefont{Aguado}},
  \bibinfo{author}{\bibfnamefont{S.}~\bibnamefont{Tarucha}}, \bibnamefont{and}
  \bibinfo{author}{\bibfnamefont{L.~P.} \bibnamefont{Kouwenhoven}},
  \bibinfo{journal}{Science} \textbf{\bibinfo{volume}{282}},
  \bibinfo{pages}{932} (\bibinfo{year}{1998}).

\bibitem[{\citenamefont{Brandes}(2005)}]{Brandes05}
\bibinfo{author}{\bibfnamefont{T.}~\bibnamefont{Brandes}},
  \bibinfo{journal}{Phys. Rep.} \textbf{\bibinfo{volume}{408}},
  \bibinfo{pages}{315} (\bibinfo{year}{2005}).

\bibitem[{sup()}]{supp}
\bibinfo{note}{See supplemental material for a discussion of
  Heisenberg-Langevin equations and phonon spectral density.}

\bibitem[{foo()}]{footnote}
\bibinfo{note}{Parenthesis refer to 95 \% confidence intervals.}

\bibitem[{\citenamefont{Scully and Zubairy}(1997)}]{QuantumOpticsBook}
\bibinfo{author}{\bibfnamefont{M.~O.} \bibnamefont{Scully}} \bibnamefont{and}
  \bibinfo{author}{\bibfnamefont{S.}~\bibnamefont{Zubairy}},
  \emph{\bibinfo{title}{Quantum Optics}} (\bibinfo{publisher}{Cambridge
  University Press}, \bibinfo{year}{1997}).

\bibitem[{\citenamefont{Weber et~al.}(2009)\citenamefont{Weber, Lindwall, and
  Wacker}}]{Weber09}
\bibinfo{author}{\bibfnamefont{C.}~\bibnamefont{Weber}},
  \bibinfo{author}{\bibfnamefont{G.}~\bibnamefont{Lindwall}}, \bibnamefont{and}
  \bibinfo{author}{\bibfnamefont{A.}~\bibnamefont{Wacker}},
  \bibinfo{journal}{Phys. Stat. Sol. B} \textbf{\bibinfo{volume}{246}},
  \bibinfo{pages}{337} (\bibinfo{year}{2009}).

\bibitem[{\citenamefont{Mahan}(2000)}]{Mahan}
\bibinfo{author}{\bibfnamefont{G.~D.} \bibnamefont{Mahan}},
  \emph{\bibinfo{title}{Many-Particle Physics}} (\bibinfo{publisher}{Plenum},
  \bibinfo{year}{2000}).

\end{thebibliography}

\setcounter{equation}{0}
\renewcommand{\theequation}{S\arabic{equation}}
\setcounter{figure}{0}
\renewcommand{\thefigure}{S\arabic{figure}}

\appendix
\section{Supplemental Material}
\emph{Heisenberg-Langevin Equations} --
We give an elementary derivation of the Heisenberg-Langevin equations used in the manuscript, which follows standard treatments found in, e.g., Refs.\ \cite{QuantumOpticsBook,MeystreBook}.  Neglecting the leads, we break the total Hamiltonian following the polaron transformation into three separate terms describing the DQD-cavity system, the phonon and photon reservoirs, and the coupling between the system and the reservoirs ($\hbar=1$)  
\begin{align}
H&=H_{S}+ H_{R}+H_{SR}, \\
H_S &= \frac{\omega_d}{2} \sigma_z +\omega_c a^\dagger a+ g_c \sin \theta (a^\dagger \sigma_-+ h.c.), \\
H_R&= \sum_{q,\nu} \omega_\nu(q) a^\dagger_{q\nu}a_{q\nu} + \sum_n \omega_n b^\dagger_n b_n, \\
H_{SR}& =-R_c^\dagger\, a - R_d^\dagger\, \sigma_- - R_e^\dagger\, a^\dagger \sigma_- - R_a^\dagger\, a \sigma_- - h.c.
\end{align}
where we introduced a bath of modes that couple to the cavity field $b_n$ and neglected counter-rotating terms and higher order terms in $(g_c,\lambda_{q,\nu})$ in $H_S$ and $H_R$.  Excluding $R_c$, we implicitly derived expressions for the reservoir operators in the main text.  Writing them out explicitly
\begin{align}
R_c &= \sum_n \tau_n b_n,\\
R_d &=-\frac{2 t_c}{\omega_d}  \sum_{q,\nu} \lambda_\nu(q) a_{q\nu}, \\
R_e&= \frac{4 t_c \epsilon}{\omega_d^2} \sum_{q,\nu} \frac{ g_c \lambda_\nu(q)}{\omega_c \omega_\nu(q)} (\omega_\nu(q)+\omega_c) a_{q\nu}, \\
R_a&=\frac{4 t_c \epsilon}{\omega_d^2} \sum_{q,\nu} \frac{ g_c \lambda_\nu(q)}{\omega_c \omega_\nu(q)} (\omega_\nu(q)-\omega_c) a_{q\nu}.
\end{align}
In a frame rotating at frequency $\omega$, we can formally express the reservoir operators in terms of the system operators:
\begin{align}
R_c(t)&=\sum_n \tau_n b_n(0) e^{-i (\omega_n - \omega)t} \nonumber \\
&+ i \int_0^t dt' \sum_n \abs{\tau_n}^2 e^{-i(\omega_n-\omega)t'} a(t'),  \label{eq:Rc}
\end{align}
and similarly for the other $R_\mu$.  The key approximation is to assume that the system operators change slowly in time so that they can be removed from the integral in the second term of Eq.\ (\ref{eq:Rc}).  This allows us to write
\begin{align}
R_c(t)& \approx i \mathcal{F}_c(t)+ i \frac{\kappa}{2} a(t) , \\
\mathcal{F}_c(t)&=-i\sum_n \tau_n b_n(0) e^{-i (\omega_n - \omega)t}, \\
\kappa& = 2 \pi \sum_n \abs{\tau_n}^2 \delta(\omega-\omega_n).
\end{align}
and similarly for the other $R_\mu$.  With these expressions, the equations of motion for the system operators close and we can solve for the dynamics solely in terms of the system operators  (provided we know the correlations of the $\mathcal{F}_\mu$).  For example, the Heisenberg-Langevin equation for $\sigma_-$ takes the form
\begin{align}
\dot{\sigma}_- &= -(\Gamma+ i \Delta ) \sigma_- + i g_c \sin \theta \,a\, \sigma_z   + \sigma_z \mathcal{F}_d \\ \nonumber
&-\sigma_z a \Big(\frac{\gamma_{e}}{2} a^\dagger \sigma_-  - \mathcal{F}_{e}\Big)  - \sigma_z a^\dagger  \Big(\frac{\gamma_{a}}{2} a \sigma_- -\mathcal{F}_{a}\Big), \nonumber
\end{align}
where $\Delta = \omega_d-\omega_c$ is the detuning of the dot from the cavity (including corrections from the polaron transformation) and $\mathcal{F}_{e(a)}$ are the noise operators associated with the phonon-assisted emission(absorption) terms.  They satisfy 
$\langle{\mathcal{F}_{\mu}^\dagger(t) \mathcal{F}_{\mu}(t')} \rangle = \gamma_{\mu}^\uparrow\,  \delta_{\mu,\mu'}  \delta(t-t')$. 
Defining $\eta_\mu(t) = \int_0^t dt' e^{-(\Gamma+i \Delta)(t-t')}\mathcal{F}_\mu(t')$,  we can express $\sigma_-(t)$ to second order in $g_c$ as 
\be
\sigma_- \approx \frac{i g_c \sin \theta\, a \sigma_z}{\Gamma+i \Delta} + a \sigma_z \eta_{e}(t)+ a^\dagger \sigma_z \eta_{a}(t)+ \sigma_z \eta_d(t).
\ee
Inserting this into the Heisenberg-Langevin equations for $a$ and $\sigma_z$, gives the mean field equations for the field amplitude $\alpha = \mean{a}$ and the population in the upper state $u= \mean{\ket{+}\bra{+}}$ shown in the main text.

\emph{Phonon Spectral Density} --
Here we discuss the frequency dependence of the phonon spectral density, including higher order modes and coupling to substrate phonons.
For the nanowire phonons, we can express the matrix elements for the electron-phonon interaction as \cite{Weber09,Weber10}
\begin{align}
\lambda_{q\nu} &= \bra{L} V_{q\nu} \ket{L} - \bra{R} V_{q\nu} \ket{R} \\ \nonumber
&=\int d \bm{x}  (\abs{\phi_\ell(\bm{x})}^2-\abs{\phi_r(\bm{x})}^2) V_{q\nu}(\bm{x}) e^{i q z} \\ \nonumber
&=2i\sin(qd/2) e^{-i q d/2} M_{q\nu},\\ 
M_{q\nu}&=\int d\bm{x} \abs{\phi(\bm{x})}^2  V_{q\nu}(\bm{x}) e^{i q z},
\end{align}
where $V_{q\nu}$ is the interaction potential for the phonon mode with quasimomentum $q$  along branch $\nu$, $z$ is the direction along the nanowire axis, $d$ is the spacing between the dots, and $\phi_{\ell(r)}=\phi$ are the envelope wavefunctions for the electrons in the dots (assumed to be identical).  We  approximate the envelopes by  a cylindrically symmetric Gaussian $\phi(\bm{x})= (a \sqrt{\pi})^{-1/2} \, \phi_r(r)\, e^{-z^2/2 a^2}$ with $a=25$ nm. The interaction potential can be broken up into a contribution from the deformation potential $V_{q\nu}^d$ and the piezoelectric potential $V_{q\nu}^p$ where the deformation potential is given by \cite{Mahan}
\begin{align}
V^d_{q\nu}(\bm{x}) &= D_\perp \nabla_\perp \cdot \bm{u}_{q\nu}(\bm{x})+ i q D_z\, \hat{z} \cdot  \bm{u}_{q\nu}(\bm{x}),
\end{align}
where $D_i$ are the deformation potential constants and $\bm{u}_{q\nu}$ is the displacement of the phonon mode such that the quantized displacement field is given by $\hat{\bm{u}}_\nu =N^{-1/2} \sum_q \bm{u}_{q\nu} e^{iqz}(a_{q\nu}+a_{-q\nu}^\dagger)$  ($N$ is the number of unit cells).
The piezoelectric potential is more difficult to calculate; however, for small $q$ it only depends on the  transverse variables and is independent of $q$.  Expanding $V_{q\nu}$ and performing the integral over space we find 
\be
M_{q\nu}= \sqrt{\frac{1}{2 M\, \omega_\nu(q) }}\,\, (i q \beta_\nu   +  \Xi_\nu) \, e^{-a^2 q^2/4}
\ee
where $M$ is the average mass of the unit cell and $\beta_\nu$ and $\Xi_\nu$ are deformation and piezoelectic constants, respectively, which depend on the different  phonon branches $\nu$ and the envelope wave function.   In the frequency range relevant to the experiment, a rough estimate shows that the piezoelectric term should dominate over the deformation potential term \cite{Weber10}.

The phonon spectral density for the nanowire is given by
\be
J(\omega)=4 \sum_{q,\nu} \sin^2(qd/2) \abs{M_{q\nu}}^2 \delta(\omega-\omega_{q\nu})
\ee
For a 25 nm radius nanowire, the higher energy waveguide modes of the nanowire have a gap of $\sim200~\mu$eV$/h$ so there is only a single phonon branch for energies below this cutoff \cite{Weber10}.  Furthermore, for small $q$ this mode is longitudinally polarized and has a linear dispersion $\omega_q = c_n q$, which implies
\be
\label{eqn:j1mode}
J(\omega) \approx  j_{n}\, \frac{{\sin}^2 (\omega d/2 c_n)}{\omega d/2 c_n} e^{-\omega^2 a^2/2 c_n^2}.
\ee
\\

In the data in Fig.\ 3(a) of the main text, there are broad tails in the gain for large detuning.   At such large frequencies our approximation for $J(\omega)$ including only the first order mode needs to be extended to include higher order modes.   We include the second order mode by taking a dispersion of the form $\omega_2(q) = \sqrt{\omega_{0}^2+c_n^2 q^2}$, which gives rise to the contribution
\be
J_2(\omega) \sim \Theta(\omega-\omega_{0})\, \frac{\sin^2 [q_2(\omega) d/2]}{q_2(\omega) d/2}\, e^{-a^2 q_2^2/2} 
\ee
where  $\Theta$ is the Heaviside step function and $q_2(\omega) = c_n^{-1} \sqrt{\omega^2 - \omega_{0}^2}$.
In addition, the phonons in the amorphous silicon nitride (SiN$_x$) substrate will begin to hybridize with the nanowire phonons, as was noted in Ref.\ \cite{Weber10}. This leads to a  contribution from the piezoelectric potential \cite{Brandes05}
\be
J_{sp}(\omega) \sim \omega\, e^{-a^2 \omega^2/2 c_s^2} \big(1 - \mathrm{sinc}(\omega d/c_s) \big),
\ee
which we expect to dominant over the deformation potential for the considered frequencies.
Here $c_s \approx 11000$ m/s is the longitudinal phonon speed of sound in SiN and $\mathrm{sinc}(x)=\sin(x)/x$.  Both these terms give a large contribution to the spectral density, whereas the contribution from the first order mode becomes exponentially suppressed at these frequencies.  Adding these terms  to $J(\omega)$ with $\hbar \omega_0 = 200~\mu$eV, we are able to account for the broad tails in the data in Fig.\ 3(a).  

\end{document}